\documentclass[twocolumn,aps,prl,showpacs,superscriptaddress,groupedaddress]{revtex4}
\usepackage{textcomp}
\usepackage{amssymb}
\usepackage{amsmath,amssymb}
\usepackage{graphicx,epstopdf}
\usepackage{color}
\usepackage{comment}

\definecolor{Com}{RGB}{0,0,0}

\begin{document}
\title{THz-bandwidth all-optical switching of heralded single photons}

\
\author{Connor~Kupchak}
\thanks{C.K. and J.E. contributed equally to this work}
\author{Jennifer~Erskine}
\thanks{C.K. and J.E. contributed equally to this work}
\affiliation{Department of Physics, University of Ottawa, Ottawa, Ontario, K1N 6N5, Canada}
\affiliation{National Research Council of Canada, 100 Sussex Drive, Ottawa, Ontario, K1A 0R6, Canada}
\author{Duncan~G.~England}
\email{duncan.england@nrc-cnrc.gc.ca}
\affiliation{National Research Council of Canada, 100 Sussex Drive, Ottawa, Ontario, K1A 0R6, Canada}
\author{Benjamin~J.~Sussman}
\email{benjamin.sussman@nrc-cnrc.gc.ca}
\affiliation{Department of Physics, University of Ottawa, Ottawa, Ontario, K1N 6N5, Canada}
\affiliation{National Research Council of Canada, 100 Sussex Drive, Ottawa, Ontario, K1A 0R6, Canada}

\begin{abstract}

\noindent Optically induced ultrafast switching of single photons is demonstrated by rotating the photon polarization via the Kerr effect in a commercially available single mode fiber. A switching efficiency of 97\% is achieved with a $\sim1.7$\,ps switching time, and signal-to-noise ratio of $\sim800$.  Preservation of the quantum state is confirmed by measuring no significant increase in the second-order autocorrelation function $g^{(2)}(0)$.  These values are attained with only nanojoule level pump energies that are produced by a laser oscillator with 80\,MHz repetition rate. The results highlight a simple switching device capable of both high-bandwidth operations and preservation of single-photon properties for applications in photonic quantum processing and ultrafast time-gating or switching.
\end{abstract}

\maketitle

The ability to quickly switch, gate, or re-route optical signals is a key component of a range of modern technologies including communications~\cite{Matsuo2010}, biomedical imaging~\cite{Andersson-Engels1990}, microscopy~\cite{Vicidomini2011}, spectroscopy~\cite{Matousek1999}, and quantum optics~\cite{Prevedel2007}. Driven by the development of modern lasers, device bandwidths measured in THz and duty cycles in GHz are available.  The speed of traditional electro-optical devices is no longer sufficient for many applications.  By contrast, all-optical approaches, which use a secondary light field to actively induce a switching mechanism, are capable of superior performance~\cite{Friberg1987,Almeida2004a,Nozaki2010} and many future technologies will rely on these techniques. Unlike electro-optic devices, the high intensity required for all-optical switching has the potential to introduce or generate unwanted photons into the switched channel. In some applications these noise photons will be negligible compared to the signal, but in others, such as quantum optics or microscopy where the signal is at the single-photon level, great care must be taken to avoid adding unwanted photons when switching. In this context, the development of all-optical switches that are simultaneously capable of high bandwidths, high duty cycles, and single-photon-level operation is emerging as an important challenge in a range of photonic disciplines. 

This letter introduces a nonlinear technique for high-speed switching of single photons based on polarization rotation via the optical Kerr effect in single-mode fiber. The method is straightforward to implement, is non-interferometric, requires only 10\,cm of conventional single mode fiber, and utilizes a low-power commercial femtosecond pump oscillator. Crucially, due to group velocity difference, the pump sweeps through the signal and induces switching.  The pulse sweeping allows pump powers to be kept below thresholds for parasitic nonlinear optical processes. The simplicity of the device means that its performance characteristics are limited only by material properties of the fiber such as dispersion and multi-photon nonlinear effects. Indeed, the high performance of the switch comes as a direct result of its simplicity. When using only 10\,cm of fiber, unwanted polarization rotations in the fiber can be kept to a minimum which allows the switch to function with high efficiency, and without an interferometer. The fiber also supports short ($\sim400$\,fs) pulses with minimal dispersion. Furthermore, the short fiber length, combined with the modest pump pulse energy requirements, contributes to the low noise floor of the device by minimizing parasitic noise processes such as Raman scattering and self-phase modulation. The system scores highly in all the key metrics outlined above: switching speeds near 1\,ps are achieved with a duty cycle of 80\,MHz, an efficiency in excess of 95\%, and a noise floor of $10^{-4}$ photons per switching window.

The noise properties of the device at the single photon level are benchmarked by measuring the second order autocorrelation $g^{(2)}(0)$ of heralded single photons that have been switched by the device. We measure $g^{(2)}(0)\simeq 0.01$ confirming that negligible noise is added by the switch and that the quantum properties of the light are maintained. We therefore expect the switch to find applications in numerous photonic quantum information schemes such as metropolitan scale quantum teleportation~\cite{Valivarthi2016}, schemes involving high-dimensional encodings~\cite{Hall2011}, and converting qubits between different degrees of freedom~\cite{Kupchak2017}. Away from quantum optics, high-contrast switching of photon level light fields can be extended to other high-bandwidth applications including spectroscopy~\cite{Takeda2000} and microscopy~\cite{Blake2016}. In these applications, and others like them, we expect this technique to be complimentary to existing single photon switching techniques based on nonlinear-optical loop mirrors (NOLM)~\cite{Hall2011,Hall2011a,Nowierski2016} and time-gating by sum-frequency generation (SFG)~\cite{Donohue2013,Donohue2014,Allgaier2017,MacLean2018,Rong2018} by offering faster switching than the former, and higher efficiency than the latter.

\begin{figure*}[htb]
\includegraphics[width=2\columnwidth]{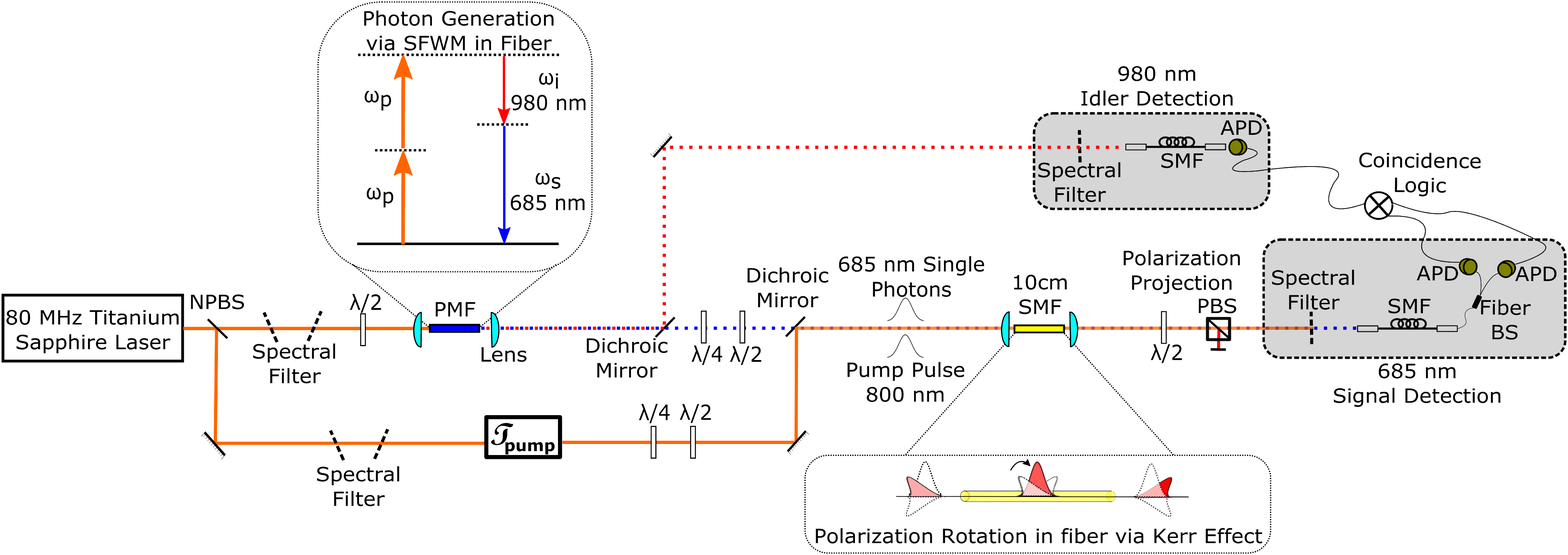} \protect\protect\caption{Schematic diagram for the experimental setup. A SFWM process occurs inside the PMF to produce signal and idler photon pairs.  Idler photons are used as a herald while signal photons undergo polarization rotation by the Kerr effect when temporally overlapped with a pump pulse (with variable delay $\tau_{\text{pump}}$) inside a SMF (shown in bottom inset). Rotated signal photons can be measured on a single APD (not shown) for coincidence detection or are split on a fiber beam splitter (BS) and sent to independent APDs (as shown) in order to measure the single photon statistics. Corresponding optical components are described in the text.}
\label{Fig:apparatus}
\end{figure*}

Heralded single photons are generated by spontaneous four wave mixing (SFWM) in a polarization maintaining fiber (PMF)~\cite{Smith2009}. After generation, the signal photons are coupled into the 10\,cm long single mode fiber (SMF) where switching is achieved by rotating the polarization of the signal photons through the optical Kerr effect. The Kerr effect is a nonlinear process whereby a pump field is used to induce birefringence in a medium with a $\chi^{(3)}$ nonlinearity.  When the medium is situated between two crossed polarizers, the incoming photons can only be transmitted when their polarization is rotated by interaction with the pump pulse. This setup is referred to as an optical Kerr shutter~\cite{Orszag1980}. The efficiency of the rotation $\eta$ can be given by~\cite{Kanbara1994}:

\begin{equation}
\eta=\sin^2\left( 2\theta \right) \sin^2\left(\frac{\Delta\phi}{2}\right),
\label{eq:efficiency}
\end{equation}
where $\theta$ is the angle between the pump and signal polarization, with the maximum switching efficiency occurring when the pump polarization is oriented at 45$^{\circ}$ with respect to the signal. The nonlinear phase shift $\Delta\phi$ is given by~\cite{Agrawal2001}:
\begin{equation}\label{eq:Bintegral}
{\color{Com} \Delta\phi = \frac{2\pi\,n_2}{\lambda_{signal}}\int_0^L I_p(T-d_wz)\,dz},
\end{equation}
where $z$ is the propagation distance along the fiber of length $L$, $n_2$ is the nonlinear refractive index, and $\lambda_{signal}$ is the wavelength of the signal photon. The intensity profile of the pump pulse $I_p$ is expressed in reduced time in the frame moving with the signal pulse $T=t-z/v_{gs}$. In a dispersive medium, the pump and signal will experience temporal walk-off given by $d_w = v_{gp}-v_{gp}$ where $v_{gp}^{-1}$ and $v_{gs}^{-1}$ are the pump and signal group velocities respectively. In many high bandwidth applications, dispersion is an inconvenience, but here it is key to the success of our scheme; by appropriately timing the pump pulse, we allow it to completely walk through the signal photon inside of the fiber. This results in a near-uniform phase-shift across the temporal profile of the signal photon. This is the most efficient way to achieve maximum switching while reducing pump energy requirements in order to minimize parasitic noise processes.

An experimental schematic is shown in Fig.~\ref{Fig:apparatus}. Both the pump and signal beams originate from an 80\,MHz repetition rate, Ti-Sapphire laser that produces pulses of 12 nm bandwidth at a central wavelength of 800 nm. A portion of the oscillator beam is split off by a non-polarizing beam splitter (NPBS) in order to pump the photon pair source. Spectral filters are used to control the bandwidth and then $\sim$33\,mW of pump power is coupled through a bow-tie style, 2.5\,cm long, PMF (Fibercore HB800)~\cite{Erskine2018}. The photon pair source generates signal and idler pairs at wavelengths of 685 and 980\,nm respectively through SFWM (see top inset in Fig.~\ref{Fig:apparatus}). After the photon source, the signal and idler photons are separated on a dichroic mirror. The idler channel is subsequently coupled to an avalanche photodiode~(APD) using a single-mode fiber (SMF) and serves as a herald for our photon counts. 

The remaining oscillator beam serves as our pump pulse.  In order to limit any noise photons generated due to self phase modulation, the pump pulse duration is lengthened using a pair of bandpass filters such that $\Delta\lambda_{\text{pump}}=$2.6~nm.  Further noise reduction is achieved by chirping the pump with 5\,cm of SF69 glass. The pump pulse is temporally combined with the signal by a variable delay ($\tau_{\text{pump}}$) and spatially combined using a second dichroic mirror.  In order to attain the peak switching efficiency governed by Eq.~(\ref{eq:efficiency}), the pulse polarizations are set to horizontal and -45$^{\circ}$ (anti-diagonal) for the pump and signal respectively. Any phase changes due to transmission of the signal through the dichroic mirror is pre-compensated using a set of quarter and half-waveplates ($\lambda /4$ and $\lambda /2$) before combination.

A 10 cm long, SMF (Thorlabs S630-HP)  is used as the Kerr medium. The use of a waveguide-based system compared to a bulk Kerr medium offers high pump intensity and excellent spatial overlap over a longer optical pathlength, reducing the pump pulse energy required to rotate and switch the signal field. The signal and pump are focused together into the 3.5 $\mu$m core using a 10 mm achromatic lens. Typically we achieve fiber coupling efficiencies of 40\% and 60\% for the signal and pump beams respectively. After propagation through the SMF,  we project the signal photons on to a diagonal (switched) or anti-diagonal (unswitched) polarization using a $\lambda /2$ and polarizing beam splitter (PBS).  Spectral filters remove the pump beam before the signal photons are coupled to an APD via SMF.

\begin{figure}[htb]
\centering
\includegraphics[width=1\columnwidth]{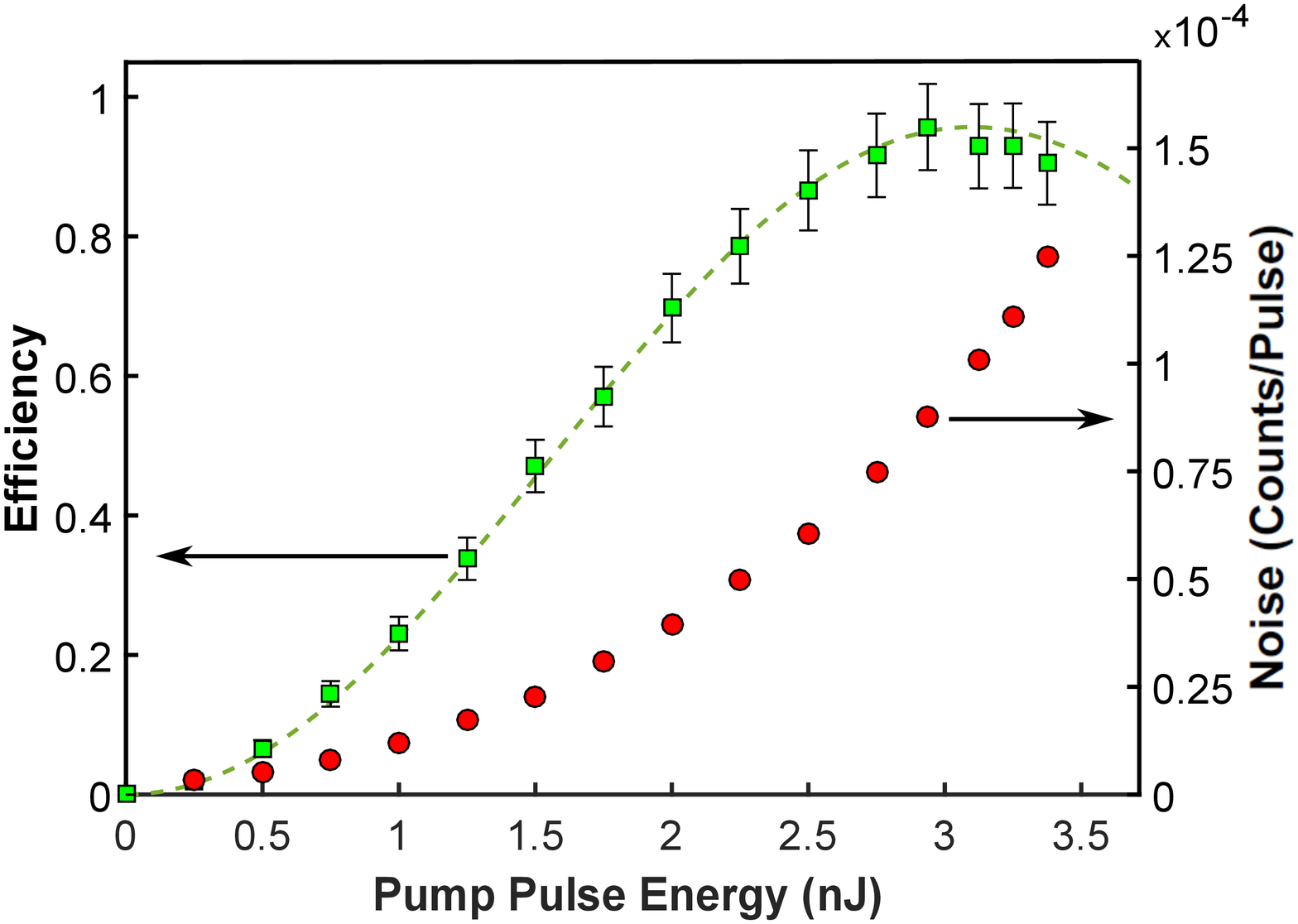}
\caption{Photon switching efficiency (left ordinate: green squares) with $\sin^2\left(\frac{\Delta\phi}{2}\right)$ fit (dashed dark green line) and the noise counts per pulse (right ordinate: red circles) as a function of pump pulse energy. The pump energy is measured at the output of the SMF.}
\label{Fig:power_scan}
\end{figure}

We characterize the technique by measuring the switching efficiency and noise statistics as a function of pump pulse energy. With zero delay between the pump and signal pulses, we adjust the energy of the pump pulse using a neutral density filter wheel (not shown in Fig.~\ref{Fig:apparatus}).  From Fig.~\ref{Fig:power_scan} we see that the efficiency continually increases to a maximum of 96\% at 3.0~nJ and follows the dependence expected from Eq.~\ref{eq:efficiency}. By blocking the signal photons we can also measure the noise characteristics. A maximum of $1.3\times10^{-4}$ noise photons per pump pulse are recorded. Here, the main noise mechanism is self-phase modulation of the pump within the Kerr medium.

To evaluate the switching response, 3.0 nJ pump pulses are temporally delayed with respect to the signal photons via a motorized stage. The switched (diagonal) polarization projection is used to measure any photons successfully rotated in the Kerr medium and are only considered when coincident with a herald photon. Likewise, the ``anti-switching" efficiency can also be evaluated by recording the absence of coincidences when the analysis optics are set to the unswitched projection.   

\begin{figure}[htb]
\centering
\includegraphics[width=.75\columnwidth]{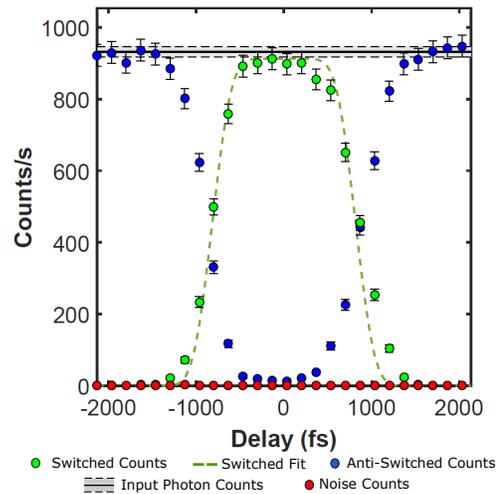}\protect\protect\caption{Polarization switching (green circles) and anti-switching (blue circles) of the 685~nm signal photons when coincident with the 980~nm idler photons. The response of the switched photons can be modeled by considering the pulse durations and the temporal walkoff between the pump and signal pulses (dark green dashed line).  Counts are shown relative to the input (non-rotated) photon count rate (solid line with shaded-grey bar denoting uncertainty) and the noise counts due to the pump (red circles). The error bars are based on Poissonian statistics.}
\label{Fig:switch_v_delay}
\end{figure}

Signal-idler coincidence counts in the switched and anti-switched polarizations are shown as a function of the pump temporal delay in Fig.~\ref{Fig:switch_v_delay}. The maximum efficiency is found to be ${\eta_{\text{switch}}=96.7\pm0.5\%}$, calculated using the mean coincidence counts in the ``flat-top" portion of the switched counts $N_{\text{switch}}$. We define the efficiency as ${\eta_{\text{switch}}=(N_{\text{switch}}-N_{\text{noise}})/N_{\text{input}}}$, where $N_{\text{noise}}$ is the mean coincidence counts measured with the signal mode blocked, and $N_{\text{input}}$ is the mean input coincidence counts. Similarly, the anti-switching efficiency calculated as ${\eta_{\text{anti-switch}}=1-(N_{\text{anti-switch}}-N_{\text{noise}})/N_{\text{input}}}$, yields a value of  ${\eta_{\text{anti-switch}}=98.0\pm0.3\%}$. The small discrepancy in these two values is from the $\sim1\%$ leakage of unswitched photons due to limitations of the polarization optics.

An important metric in any component is the signal-to-noise ratio (SNR). Here, the SNR, calculated as $\mathrm{SNR} =N_{\text{switch}}/N_{\text{noise}}$ was found to be $790\pm70$. Note that our SMF technique achieves a more than 80-fold increase in the SNR when compared to that found in a bulk crystal setup with an amplified pump~\cite{Kupchak2017}. This improvement can primarily be attributed to a better spatial overlap supplied by the fiber and substantially lower pump pulse energies.    

We numerically evaluate the B-integral in Eq.~(\ref{eq:Bintegral}) as a function of pump delay $\tau_{pump}$ by considering the temporal profile of the pump pulse and the pump-signal walk off. The pump pulse was measured by autocorrelation to be 410\,fs, and the walk off is calculated from the Sellmeier equation to be 1.6\,ps. The resulting delay-dependent nonlinear phase-shift $\Delta\phi(\tau_{pump})$ is inserted into Eq.~(\ref{eq:efficiency}) to determine the intrinsic efficiency response function of the switch. The intrinsic response function is independent of the switch input and its width determines the fastest possible switching speed. To calculate the switching efficiency of the input photons, we integrate the intrinsic response function over the duration of the signal photon, weighted by the temporal profile of the signal photon. The temporal profile of the input photons is estimated to be the convolution of the 100\,fs long pump pulse and the 380\,fs pump-signal walk off in the PMF. This total efficiency curve (green dashed line in Fig.~\ref{Fig:switch_v_delay}) has a width of $\Delta\ t_{\text{calc}}=1.63$\,ps, in good agreement with the measured duration ($t_{\text{switch}}=1.69\pm0.02$\,ps). The intrinsic switching speed is limited by the length of the fiber and by the need for the pump and signal pulse to completely walk through each other. We note that faster switching speeds could be achieved by using a shorter fiber and a shorter pump pulse, but the pump duration cannot be decreased indefinitely due to self-phase modulation and temporal broadening in the fiber.

To examine how the switch affects the non-classical properties of the single photons, we measure the heralded second order autocorrelation at zero time delay $g_{\text{switched}}^{(2)}(0)$ of the switched signal photons. For this measurement we use a Hanbury Brown-Twiss configuration~\cite{Hanbury1956} by sending the polarization rotated photons to a 50:50 fiber beam splitter (Fig.~\ref{Fig:apparatus}), with each exit port coupled to independent APDs. The $g_{\text{switched}}^{(2)}(0)$ value of the heralded SFWM source can be calculated by~\cite{Grangier1986}

\begin{equation}
g_{\text{switched}}^{(2)}(0)=\frac{P_{\text{1,2,i}}}{P_{\text{1,i}}P_{\text{2,i}}},
\end{equation}
where $P_{\text{1,2,i}}$ is the probability of a 3-fold coincidence between the idler and both signal detectors, and $P_{\text{1,i}}$ and $P_{\text{2,i}}$ are the probabilities of a two-fold coincidence between the idler and signal detector 1, and the idler and signal detector 2 respectively.

\begin{figure}[htb]
\centering
\includegraphics[width=1\columnwidth]{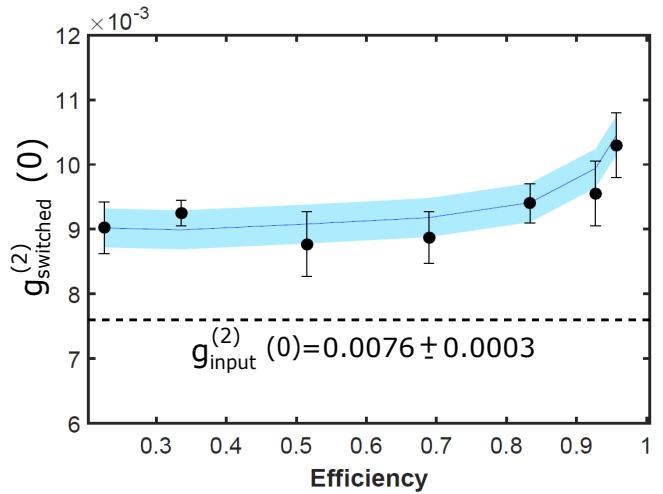}
\protect\protect\caption{Second-order correlation function $g_{\text{switched}}^{(2)}(0)$ of the heralded, switched signal photons (filled circles) compared to that of the input signal photons (dashed line). The increase in the autocorrelation value of the switched photons can be modeled by considering the noise counts (blue line with shaded region denoting the uncertainty).}
\label{Fig:g2}
\end{figure}

Fig.~\ref{Fig:g2} shows the value of the second-order autocorrelation function of the heralded, switched photons $g_{\text{switched}}^{(2)}(0)$ for increasing pump pulse energies. The input signal photons demonstrate good statistical properties, with $g_{\text{input}}^{(2)}(0)=0.0076\pm0.0003$ and the switched photons show only a modest increase the measured $g_{\text{switched}}^{(2)}(0)$. The increase in the second-order autocorrelation function can be explained by modeling the measured switched signal as an incoherent mixture of input signal and noise photons~\cite{Michelberger2015} and calculating the resulting expected autocorrelation according to

\begin{equation}
g_{\text{expected}}^{(2)}(0)=\frac{N_{\text{s,i}}^2 g_{\text{input}}^{(2)}(0)+2N_{\text{s,i}}N_{\text{noise,i}}+N_{\text{noise,i}}^2 g_{\text{noise}}^{(2)}(0)}{(N_{\text{s,i}}+N_{\text{noise,i}})^2}.
\end{equation}
Here, $g_{\text{input}}^{(2)}(0)$ is the measured heralded autocorrelation of the input photons, $N_{\text{s,i}}$ is the number of heralded signal photons, $g_{\text{noise}}^{(2)}(0)$ is the second order autocorrelation of the noise photons, and $N_{\text{noise,i}}$ is the number of heralded noise photons. The second-order autocorrelation of the noise photons was measured to be $g_{\text{noise}}^{(2)}(0)=1.07 \pm 0.05$. From Fig.~\ref{Fig:g2}, we can see that this incoherent model closely matches the recorded increase in the measured $g_{\text{switched}}^{(2)}(0)$ values.

In summary, we demonstrate single photon switching at picosecond timescales using a commercially available single mode fiber and pump oscillator. The method achieves high efficiency, high duty cycle, excellent SNR values, requires only nanojoule pump energies, and maintains  nonclassical single photon statistics. It is easy to implement and no active stabilization or complex optical components are needed. We note that the pump energy requirements are within reach of commercial fiber oscillators offering the opportunity for a completely integrated switch. {\color{Com} The} technique provides flexibility in the range of pump and signal wavelengths, {\color{Com} and can easily be adapted to provide a phase-shift instead of a polarization rotation}. We expect {\color{Com}this tool} to be applicable in a range of quantum optical processing applications such as the conversion of photonic qubits~\cite{Kupchak2017}, optical computing in a single spatial mode~\cite{Humphreys2013}, and the processing of high dimensional and hyperentangled quantum states~\cite{Hall2011, Ikuta2017, Ikuta2018}. Another clear application is in optical time-gating; the Kerr switch {\color{Com} offers an} alternative to the commonly used SFG approach~\cite{MacLean2018, Donohue2013}. {\color{Com}Lastly, it is expected that this technique will have applications that extend beyond quantum protocols by providing timing selectivity in spectroscopy~\cite{Takeda2000} and microscopy~\cite{Blake2016}.}

This work is supported by the Natural Sciences and Engineering Research Council of Canada.
The authors thank Khabat Heshami, Rune Lausten, Denis Guay, and Doug Moffatt for useful discussions and technical assistance

\section{References}

\bibliography{OKS}

\end{document}